\RequirePackage{fix-cm}
\documentclass[twocolumn]{svjour3}          
\smartqed  
\usepackage{graphicx}
\usepackage{microtype}
\usepackage{subfigure}
\usepackage{booktabs} 
\usepackage{amssymb}
\usepackage{amsmath}
\usepackage{makecell}
\usepackage{multicol, multirow}
\usepackage{xcolor}

\usepackage{amsthm}  

\newtheorem{hypothesis}{Hypothesis}

\usepackage{hyperref}
\hypersetup{breaklinks=true}
\usepackage{url}

\usepackage{array}
\newcolumntype{L}[1]{>{\raggedright\let\newline\\\arraybackslash\hspace{0pt}}m{#1}}
\newcolumntype{C}[1]{>{\centering\let\newline\\\arraybackslash\hspace{0pt}}m{#1}}
\newcolumntype{R}[1]{>{\raggedleft\let\newline\\\arraybackslash\hspace{0pt}}m{#1}}

\usepackage{epsfig}
\usepackage{color}

\usepackage{amscd}

\def\({\left(}
\def\){\right)}

\definecolor{sam}{named}{black}
\definecolor{hadi}{named}{black}

\begin{document}

\title{Learning with Mandelbrot and Julia
}
\subtitle{Classification of fractal sets using machine learning}


\author{{V.R. Tjahjono} \and {S.F. Feng} \and {E.R.M. Putri} \and {H. Susanto}}


\institute{
V.R. Tjahjono \at Faculty of Mathematics and Natural Sciences,\\Institut Teknologi Bandung, Bandung 40132, Indonesia\\
\email{20821026@mahasiswa.itb.ac.id}
\and
S.F.\ Feng \at  \textcolor{sam}{Sorbonne University Abu Dhabi, \\SAFIR,\\Sorbonne Center for Artificial Intelligence,}\\ Abu Dhabi, United Arab Emirates\\
\email{samuel.feng@sorbonne.ae}
\and
E.R.M.\ Putri \at {Department of Mathematics,\\ Faculty of Mathematics and Natural Sciences,\\ Institut Teknologi Sepuluh Nopember, Sukolilo,\\ Surabaya 60111, Indonesia}\\ 
\email{endahrmp@matematika.its.ac.id}
\and 
H.\ Susanto \at {Department of Mathematics, Khalifa University,\\ PO Box 127788, Abu Dhabi, United Arab Emirates}\\ \email{hadi.susanto@yandex.com}           
}

\date{Received: date / Accepted: date}

\maketitle

\begin{abstract}
Recent developments in applied mathematics increasingly employ machine learning (ML)—particularly supervised learning—to accelerate numerical computations, such as solving nonlinear partial differential equations. In this work, we extend such techniques to objects of a more theoretical nature: the classification and structural analysis of fractal sets. Focusing on the Mandelbrot and Julia sets as principal examples, we demonstrate that supervised learning methods—including Classification and Regression Trees (CART), K-Nearest Neighbors (KNN), Multilayer Perceptrons (MLP), and Recurrent Neural Networks using both Long Short-Term Memory (LSTM) and Bidirectional LSTM (BiLSTM)\textcolor{sam}{, Random Forests (RF), and Convolutional Neural Networks (CNN)}—can classify fractal points with significantly higher predictive accuracy and substantially lower computational cost than traditional numerical approaches, such as the thresholding technique. These improvements are consistent across a range of models and evaluation metrics. Notably, KNN \textcolor{sam}{and RF exhibit} the best overall performance, and comparative analyses \textcolor{sam}{between models} (e.g., KNN vs.\ LSTM) suggest the presence of novel regularity properties in these mathematical structures. Collectively, our findings indicate that ML not only enhances classification efficiency but also offers promising avenues for generating new insights, intuitions, and conjectures within pure mathematics.

\end{abstract}

\section{Introduction}
\label{sec:introduction}

Fractals are geometric objects—such as curves, surfaces, and volumes—characterized by intricate, often angular structures that exhibit self-similarity across scales. They frequently appear in the morphology of natural phenomena and are encountered in various physical systems during experimental studies \cite{gouyet1996physics}. The term \emph{fractal} was introduced by Beno\'it Mandelbrot in the 1970s, and since then, fractals have garnered significant attention both for their visual elegance and for their utility in accurately modeling complex physical structures and multiscale processes \cite{mandelbrot1982fractal,kunze2011fractal}.

Fractals often arise as geometric manifestations of dynamical systems in appropriate phase spaces. A central concern in dynamical systems theory is the asymptotic behavior of orbits, particularly in the presence of nontrivial recurrence \cite{katok1995introduction}. Discrete iterated maps on the complex plane—equivalent to a special class of two-dimensional real maps satisfying the Cauchy-Riemann equations—partition the space into two distinct sets: those with bounded orbits and those with unbounded orbits. Focusing on quadratic maps as a prototypical example yields the Mandelbrot and Julia sets, which are defined by the boundedness of orbits and are notable for their intricate fractal boundaries \cite{devaney1993first}.

Recent advances in the study of fractal sets include, for instance, the work of Danca and Fe{\v{c}}kan \cite{danca2023mandelbrot}, who introduced fractional Mandelbrot and Julia sets by employing discrete fractional calculus as a natural extension of traditional difference calculus. Their analysis covers several key properties, including dynamical behavior, boundedness, and symmetry. Kawahira \cite{kawahira2020zalcman} revisited Tan's celebrated theorem \cite{lei1990similarity}, establishing a precise correspondence between the Mandelbrot and Julia sets at Misiurewicz points. In another direction, Dobbs \cite{dobbs2023hausdorff} investigated the Hausdorff dimension of quadratic Julia sets and derived explicit bounds for this typically discontinuous quantity near the tip of the Mandelbrot set. Additionally, Jaksztas \cite{jaksztas2023directional} examined the directional derivative of the Hausdorff dimension, providing an asymptotic formula that characterizes its variation.

In this paper, we argue that machine learning (ML) can—and indeed should—be employed as a tool for studying fractal sets. While this proposal is not entirely novel, as mutual interactions between the two domains already exist, we aim to refine and extend this interplay within a more formal mathematical context. Broadly speaking, the interaction between fractal geometry and ML can be categorized into two main directions: the application of ML algorithms to analyze fractal data, and the use of fractal structures to inform the design of novel ML models.

In the first category, studies typically focus on the computation and utilization of fractal dimensions in applied contexts, such as malignancy prediction in tumors \cite{chan2016automatic}, segmentation of skin lesion borders \cite{ali2020machine}, and landslide susceptibility mapping \cite{hu2020machine}. In the second category, recent work includes the development of fractal-based predictive models with finite memory, such as fractal prediction machines \cite{tino2001predicting} that learn statistical regularities in discrete sequences of data using geometric representations of sequence contexts based on fractal transformations, and fractal image compression techniques using schema-based genetic algorithms that leverage the self-similarity of natural images \cite{wu2007schema}. \textcolor{hadi}{Kataoka et al.\ \cite{kataoka2020pre} introduce a framework that pre-trains Convolutional Neural Networks (CNNs) using fractal-generated images and labels derived from natural laws, rather than natural images. It achieves competitive or superior performance compared to ImageNet and Places pre-trained models in certain settings, offering distinctive feature representations in learned convolutional and attention maps. The authors of \cite{anderson2022improving} extend prior work on fractal-based pre-training by introducing a dynamic fractal image dataset and a novel multi-instance prediction task, enabling efficient pre-training of deep neural networks without natural images. Their approach achieves up to 98.1\% of the performance of ImageNet pre-trained models while avoiding issues of label noise, dataset bias, and data curation cost. Tu et al.\ \cite{tu2023learning} propose a novel method to generate fractal images resembling arbitrary target images by learning fractal parameters through gradient descent. Their approach demonstrates high visual fidelity and flexibility across loss functions, offering potential for applications in visual representation learning and scientific analysis.}

In contrast to these application-driven approaches, our investigation is motivated by a pure mathematics perspective \cite{katok1995introduction}. As an initial step, we focus on the Mandelbrot and Julia sets, which have emerged as canonical examples in the study of fractals.


The application of ML to Mandelbrot sets has been explored in several prior—but largely unpublished—studies (see, e.g., \cite{mastersthesis,chatel}). However, these efforts primarily use Mandelbrot sets as a testbed for algorithmic demonstrations or for pedagogical illustration, rather than as objects of systematic mathematical inquiry. In contrast, the present work undertakes a structured investigation of these sets, with our primary contributions summarized as follows:
\begin{itemize}
    \item We design and conduct a series of computational experiments to evaluate the effectiveness of supervised learning techniques in the classification of fractal sets, focusing on the Mandelbrot and Julia sets as canonical examples.
    
    \item We find that accurate classification of asymptotic boundedness can be achieved using only a surprisingly small \textcolor{sam}{($\leq 4$)} number of iterates from the underlying dynamical system.
    
    \item We show that a broad range of standard supervised learning models—specifically CART, KNN, MLP, LSTM, \textcolor{sam}{BiLSTM, RF, and CNN}—consistently outperform the traditional numerical thresholding techniques widely employed in fractal analysis.
    
    \item We conduct comparative performance analyses of these models, which in turn motivate novel \textcolor{sam}{directions for future investigation} about previously unexamined regularities in the structure of fractal sets.
    
    \item \textcolor{sam}{We produce a novel conjecture on the existence of a classification function that maps a finite number of early orbit iterates to the boundedness of fractal points near the boundary, suggesting an underlying geometric regularity in such sequences.}
    
    \item We release all associated code to enable reproducibility and to facilitate further exploration of other fractal sets using similar methodologies.
\end{itemize}

Taken together, our findings suggest that ML offers a valuable and underutilized framework for the mathematical study of fractals. Beyond improving upon existing algorithmic methods, supervised learning models—through their comparative performance and structural behavior—may reveal new insights into the geometry and dynamics of these extensively studied mathematical objects.

\section{Background and Notation}

In this section, we setup our basic notations and describe the fractal objects of our study.

\subsection{The Mandelbrot Set}
\label{sec:mandelbrot}
Mandelbrot set was firstly uncovered in \cite{mandelbrot1982fractal}. Since then, the Mandelbrot set has stimulated extensive research topics in mathematics and has also been the foundation for an uncountable quantity of graphics outlines, hardware demos, and web pages. 
\par
\begin{figure}[ht]
\vskip 0.2in
\begin{center}
\centerline{\includegraphics[width=\columnwidth]{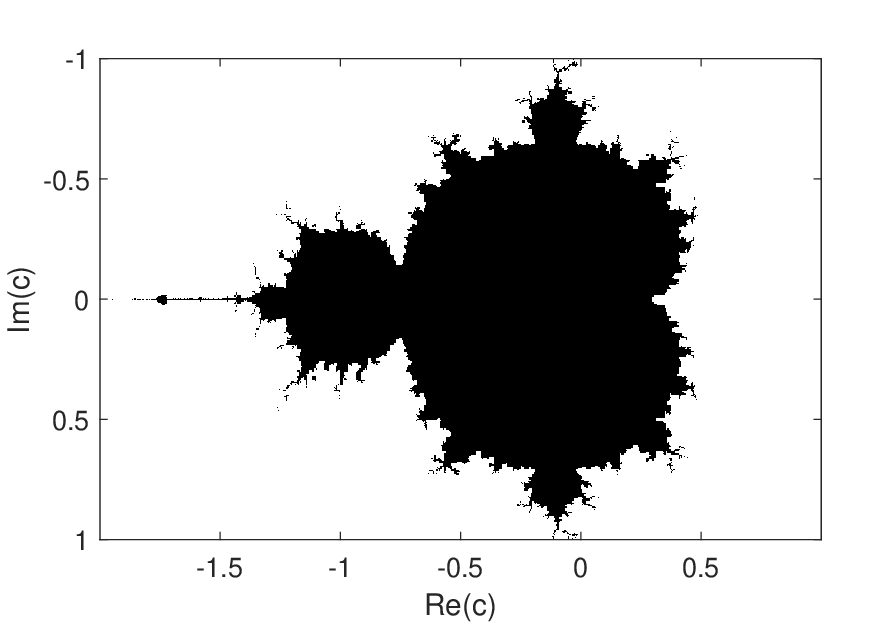}}
\caption{Mandelbrot's fractal set in the complex plane.}
\label{fig:mandelbrot}
\end{center}
\vskip -0.2in
\end{figure}

 The idea of \textcolor{sam}{the} Mandelbrot set \textcolor{sam}{is based} on the complex quadratic \textcolor{sam}{recurrence relation} \begin{equation}
    z_{k+1}=z_k^2+c,\label{qm}
\end{equation} which \textcolor{sam}{is} topologically conjugate to a quadratic polynomial $Q_c(z)=z^2+c$ for $c\in\mathbb{C}$. The Mandelbrot set, $\mathcal{M}$, is defined as the collection of all $c$ for which the orbit $\{z_k\}$ of the point $0$, which is the only critical point of $Q_c$ (i.e., the only point that satisfies $Q'_c(z)=0$) remains bounded \cite{mandelbrot1983quadratic}, that is, $$\mathcal{M}=\lbrace c\in\mathbb{C}\big|\lim_{k\to\infty} |Q_c^k(0)|\nrightarrow\infty
\rbrace,$$ where $Q_c^0(z)=z$, $Q_c^{k+1}(z)=Q_c(Q^k_c(z))$ for $k = 0,1,\ldots$. Throughout this paper, we will often use the shorthand $z_k:= Q_c^{k}(z)$ to denote the $k$th nonzero term of the Mandelbrot orbit, so that $c = z_1$ is the first nonzero term of the orbit. \textcolor{sam}{A visualization of the Mandelbrot set is provided in Fig.\ \ref{fig:mandelbrot}.}

\subsection{The Julia Set}
\label{sec:julia}
\begin{figure}[ht]
\begin{center}
\includegraphics[width=\columnwidth]{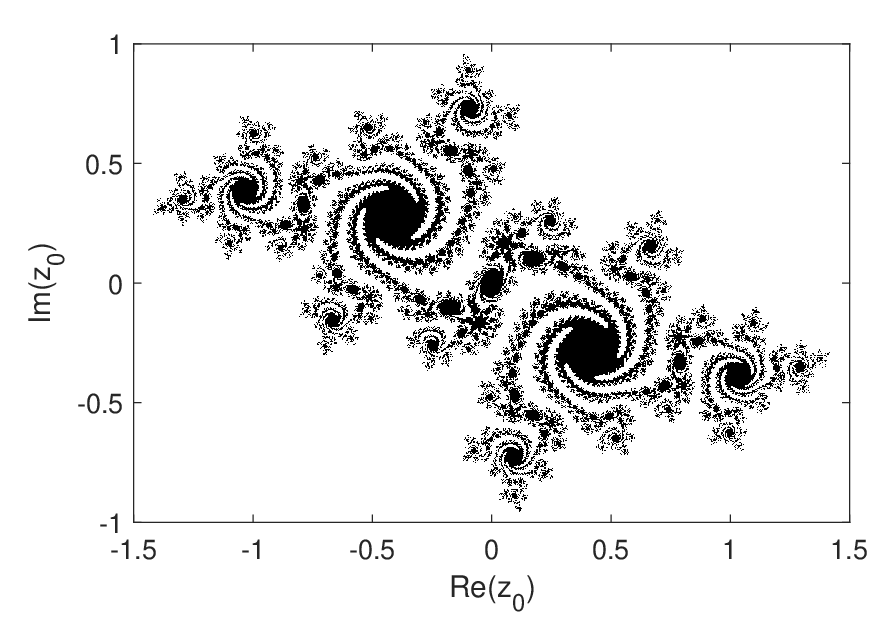}
\vskip -0.2in
\caption{The fractal structure of the Julia set on the complex plane for $c=-0.5125+0.5219i$.}
\label{fig:julia}
\end{center}
\end{figure}

Julia set, $\mathcal{J}_c$, is defined as a collection of complex numbers $z_0\in \mathbb{C}$ for a fixed $c\in\mathbb{C}$ for which the orbit initiated from the point $z_0$ from the quadratic map \eqref{qm} remains bounded \cite{devaney1993first}, that is, 
$$\mathcal{J}_c=\lbrace z_0\in\mathbb{C}\big|\lim_{k\to\infty} |Q_c^k(z_0)|\nrightarrow\infty 
\rbrace,$$ where $Q_c^{k}(z)$ is defined similarly as before. Julia sets therefore complement the Mandelbrot set. To be precise, for values of $c$ that are inside the Mandelbrot set, one will obtain connected Julia sets, i.e., all the black regions in the Julia sets are connected \cite{blanchard1986disconnected}.  
\textcolor{sam}{A visualization of the Julia set is provided in Fig.\ \ref{fig:julia}.}

\section{Other Related Work}
\label{sec:relatedwork}

Recent research on the application of ML to dynamical systems has integrated supervised and deep learning techniques for forecasting and classifying time series data. Liu and Hauskrecht propose a hierarchical dynamical system framework that employs ML and data mining methods to effectively handle irregularly sampled time series, enhancing predictive performance \cite{liu2015clinical}. Castillo and Melin develop an intelligent system that learns from a dynamical model of financial data to streamline and accelerate the forecasting process \cite{castillo1995intelligent}.

\textcolor{sam}{CNNs have demonstrated strong performance in classifying time series data derived from chaotic systems \cite{boulle2020classification}. However, in domains characterized by long-range dependencies or irregular temporal structures, architectures such as Recurrent Neural Networks (RNNs), Long Short-Term Memory (LSTM) networks, and attention-based models often yield superior results \cite{ismail2019deep,ismail2020inceptiontime}. These findings underscore that the optimal choice of model architecture is highly context-dependent.}

\textcolor{hadi}{ML has also been employed to solve dynamical systems governed by evolutionary partial differential equations (PDEs) in high-dimensional settings, thereby mitigating the curse of dimensionality that limits traditional deterministic methods \cite{grohs2023proof,han2018solving,michoski2020solving,putri2024deep}. Another significant development at the intersection of nonlinear dynamics and ML is the introduction of physics-informed neural networks (PINNs). These networks are trained to perform supervised learning tasks while satisfying physical laws expressed through nonlinear PDEs \cite{raissi2018deep,raissi2018hidden,raissi2019physics}. Unlike traditional numerical approaches, which require fine discretization and are computationally demanding, PINNs utilize automatic differentiation and variational principles to approximate PDE solutions with less data and computational overhead.}

\textcolor{sam}{PINNs have been applied to analyze bifurcation diagrams in both continuous and discrete dynamical systems and to solve associated linear eigenvalue problems \cite{shahab2024neural,shahab2025corrigendum,shahab2025neural}. Recent efforts to enhance their accuracy and scalability include the incorporation of neural tangent kernels \cite{wang2021understanding}, domain decomposition strategies \cite{jagtap2020extended}, multi-fidelity models \cite{meng2020composite}, and transfer learning methods \cite{daneker2024transfer}. Robinson et al.\ \cite{robinson2022physics} demonstrate that embedding partial physical knowledge within intermediate layers of deep neural networks improves accuracy, reduces uncertainty, and accelerates convergence when modeling nonlinear systems with invariant fractal structures. Zhang and Gilpin \cite{zhang2024zero} show that foundation models pre-trained on diverse time-series datasets can achieve competitive zero-shot forecasting on chaotic systems, preserving geometric and statistical properties such as fractal dimensions—even beyond the regime where traditional methods fail.}

\textcolor{sam}{The Sparse Identification of Nonlinear Dynamical Systems (SINDy) framework, introduced by Brunton, Proctor, and Kutz \cite{brunton2016discovering}, offers a data-driven methodology for uncovering governing equations from time-series observations. SINDy models system dynamics as a sparse linear combination of candidate functions selected from a predefined library (e.g., polynomials, trigonometric terms), using regression techniques such as sequential threshold least squares or LASSO. This approach aligns with the principle that many complex behaviors arise from simple, interpretable laws. The framework has been extended to discover PDEs from data \cite{champion2019data}. Gilpin \cite{gilpin2021chaos} contributes a large, extensible database of chaotic dynamical systems annotated with time series and mathematical properties (e.g., fractal dimensions), enabling systematic benchmarking and advancing data-driven modeling applications.}

\section{Experimental Design and Implementation}
In this section, we present the numerical experiments that underpin the main results of this paper. Simulated data generated from mathematical structures offer a key experimental advantage: they allow precise control over subtle aspects of the setup, thereby facilitating reproducibility and interpretability.

\subsection{Methods for Classification}
\label{sec:methods}
A classical technique used by fractal researchers to classify fractal points is the threshold method, which we refer to as {THRESH} \cite{mandelbrot1983quadratic}. To numerically compute $\mathcal{M}$, we leverage the fact that the orbit escapes to infinity if and only if $|z_n| < |z_{n+1}|$ for all $n > N_0$, for some $N_0 > 0$. Applying the triangle inequality—i.e., $|z_n|^2 - |c| \leq |z_n^2 + c| \leq |z_n|^2 + |c|$—we find that a sufficient condition for escape is $|z_n| > r_w := (1 + \sqrt{1 + 4|c|}) / 2$. This critical value, $r_w$, is known as the radius of the ``whirlpool circle," which lies outside the domain of attraction of $|z| = \infty$ \cite{mandelbrot1983quadratic}.

A simpler sufficient condition for divergence is given by $|z_n| > 2$ and $|z_n| > |c|$. The proof follows directly: under these conditions, we have
\[
|z_{n+1}|^2 > |z_n|^2 - |c| > |z_n|^2 - |z_n| = |z_n|(|z_n| - 1) > |z_n|.
\]
Since $|z_1| = |c|$, the orbit will be unbounded if $|c| > 2$. For simplicity, we adopt the fixed threshold $r_w = 2$ throughout this work, noting that in general, $r_w$ depends on the value of $c$.

In addition to the THRESH method, we evaluate {seven} standard classification techniques from the ML toolkit \textcolor{sam}{(see Table \ref{tab:models})}: Classification and Regression Tree (CART), \textcolor{sam}{Random Forests (RF), one-dimensional Convolutional Neural Network (CNN)}, K-Nearest Neighbors (KNN), Multilayer Perceptron (MLP), Recurrent Neural Network (RNN) with Long Short-Term Memory (LSTM), and RNN with Bidirectional LSTM (BiLSTM). Details on hyperparameters, model architectures, and implementation are provided in Section~\ref{sec:implementation} and Appendix~\ref{appA}.


\subsection{Simulation, Training, and Testing}
\label{sec:boundary}

We consider three distinct domains: the Mandelbrot set over $[-2, 1] \times [-1, 1]$, \textcolor{sam}{a Julia set with parameter $c = -0.5125 + 0.52129i$ over $[-1.5, 1.5] \times [-1, 1]$, and a region near the boundary of the Mandelbrot set}.

To sample points near the Mandelbrot boundary, we numerically approximate the boundary by solving the equation $|Q_c^{k}(0)| - r_w = 0$ for a large integer $k$ (typically $k = 100$). At each boundary point $c$, we define a square box of side length $0.07$ centered at that point. Sample points within each box are then drawn uniformly.

For each of the three domains considered, we uniformly sampled points to generate training sets of 10,000 samples. The labels (bounded vs. unbounded) were determined using the classical threshold algorithm, which was run for 100 iterations. A point was labeled as unbounded if any of the first 100 iterations of its orbit had a magnitude greater than \textcolor{sam}{$r_w = 2$} (see Section~\ref{sec:methods}). The input features consisted of the real and imaginary parts of the points: $\operatorname{Re}(c), \operatorname{Im}(c), \operatorname{Re}(z_2), \operatorname{Im}(z_2), \dots, \operatorname{Re}(z_I), \operatorname{Im}(z_I)$ for integer values of $I \geq 1$.\footnote{For the Mandelbrot set, the input variables are as described. For the Julia set, the input variables are $\operatorname{Re}(z_0), \operatorname{Im}(z_0), \operatorname{Re}(z_1), \operatorname{Im}(z_1), \dots, \operatorname{Re}(z_{I-1}), \operatorname{Im}(z_{I-1})$.} The typical value for $I$ was $4$, resulting in an input feature space of dimension \textcolor{sam}{$8$}, as we are dealing with complex numbers. After fitting each model to the training data, we evaluated its performance on a test set of 1,000,000 points, which were uniformly sampled from each domain, with labels again determined using the same \textcolor{sam}{long orbit} classical algorithm. This procedure was repeated 10 times, each time with a new independent training set, yielding an estimate of both the classification accuracy and the distribution of accuracy for each model.

\subsection{Implementation and Code}
\label{sec:implementation}

All simulations were performed on a personal computer with the following specifications: Intel i5-8250U, Windows 10 64-bit, and 12GB of RAM. The simulation, modeling, and analysis were carried out using MATLAB (R2019b Academic License) and Python. The MLP was implemented in Python using \texttt{scikit-learn v0.23.2}. KNN and CART were implemented using the Statistics and Machine Learning Toolbox in MATLAB, while LSTM and BiLSTM models were implemented using MATLAB's Deep Learning Toolbox. Violin plots and related analyses were conducted in Python using \texttt{Matplotlib v3.1.3}.

Each neural network was trained with a learning rate of 0.005, which was reduced by a factor of 0.2 after every 100 epochs. The total number of iterations for training each neural network was set to 500. For the MLP, the number of neurons in the hidden layer was determined using the following formula:
$$
Ne = \left[\dfrac{2}{3}\left(In + Out\right)\right]
$$
where $Ne$ is the number of neurons in the hidden layer, $In$ is the number of input features (i.e., attributes), and $Out$ is the number of outputs (i.e., targets or labels). The rounding operator $\left[\,\,.\,\,\right]$ is used to ensure the result is an integer. For $I=4$, this yields $Ne = \frac{2}{3} \times (9 + 1) = 7$ neurons in the hidden layer. For both LSTM and BiLSTM models, 50 hidden neurons were used in each layer.

{The RF model was configured with an ensemble of 100 decision trees. Key parameters included setting the method to \texttt{classification}, which is appropriate for categorical outcomes, and enabling Out-of-Bag (OOB) error estimation.}

The 1-D CNN classifier was implemented using a convolutional layer with ReLU activation to extract temporal features from the input data. This was followed by dropout regularization and a global max pooling layer, which reduces the dimensionality while retaining important features. The extracted features were then passed through two fully connected layers: the first with ReLU activation and the second with a sigmoid activation function to produce the final binary output.

We note that for sequential models (LSTM, BiLSTM, and CNN), the input orbits are treated as ordered sequences, as these models are designed to process sequential data. In contrast, for the other models, the input orbits are treated as unordered flat vectors. Table~\ref{tab:models} lists all of the models considered in this study, and further details on model architectures and parameters can be found in Table~\ref{tab:parameters} in Appendix~\ref{appA}. 

A GitHub repository containing all implementation details, enabling reproduction of our results, is available at \cite{fractalpaper}. 

    
    \begin{table}[htbp!]
        \centering
        \vskip 0.15in
        \begin{tabular}{cc}
            \hline
            \textbf{\underline{Model Name}} & \textbf{\underline{Abbreviation}} \\
            Simple Threshold $|z_n|>2$ & THRESH\\
            Classification and Regression Tree & CART\\
            $k$ nearest neighbor, $k=3$ & KNN\\
            Multilayer Perceptron & MLP\\
            Long Short-Term Memory Neural Net & LSTM\\
            Bidirectional LSTM & BiLSTM\\
            Random Forest & RF\\
            1-D Convolutional Neural Network & CNN\\
            \hline
            
        \end{tabular}
    
        \caption{\textcolor{sam}{Models considered in this paper and their abbreviations. Model architecture and parameter details can be found in Table \ref{tab:parameters} in Appendix \ref{appA}.}}
        \label{tab:models}
    \end{table}

\section{Results}

In this section, we present the results of our numerical experiments.

\subsection{Mandelbrot Set}
\label{sec:results-mandelbrot}
\begin{figure}[ht!]
\vskip 0.2in
\begin{center}
\centerline{\includegraphics[width=\columnwidth]{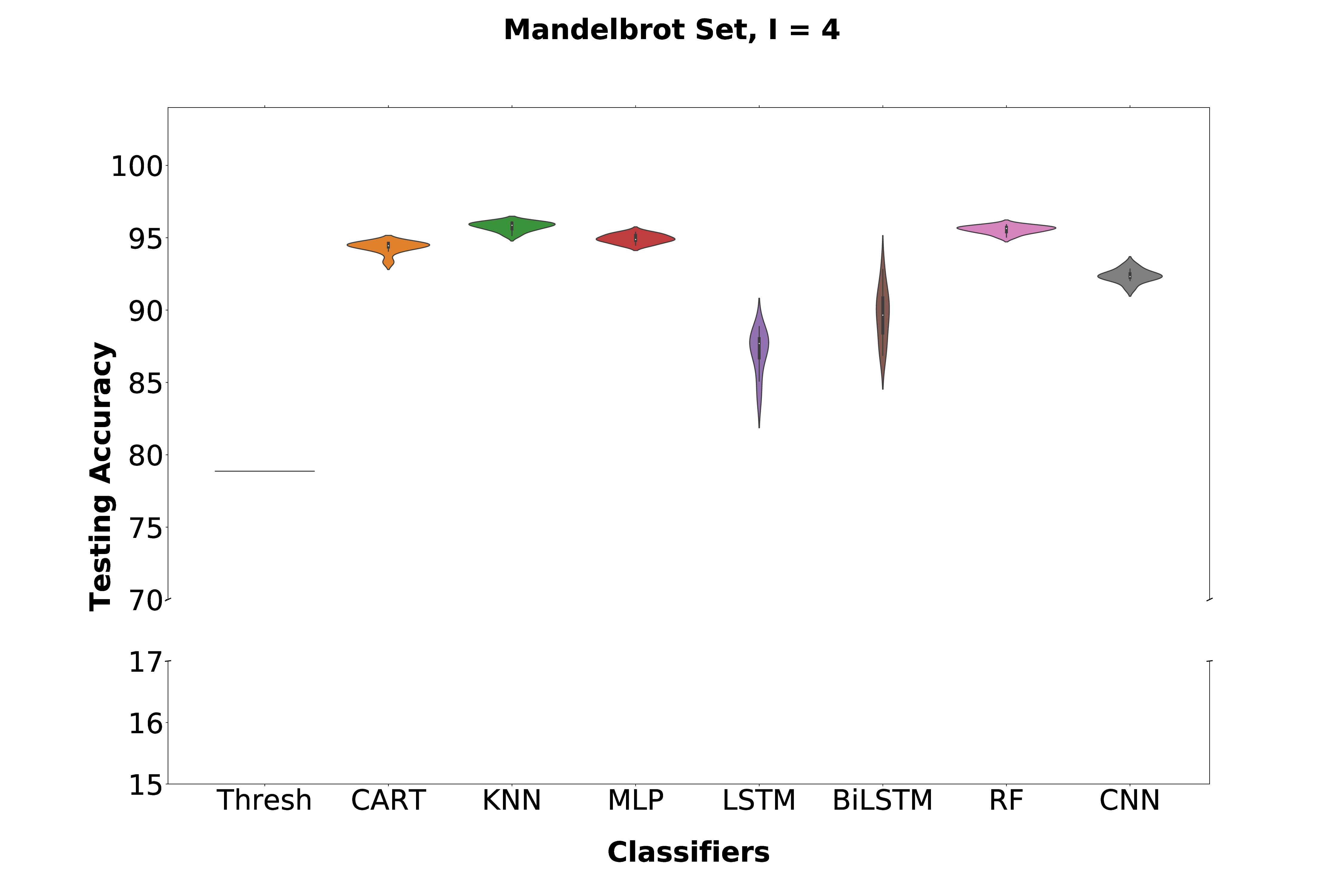}}
\centerline{\includegraphics[width=\columnwidth]{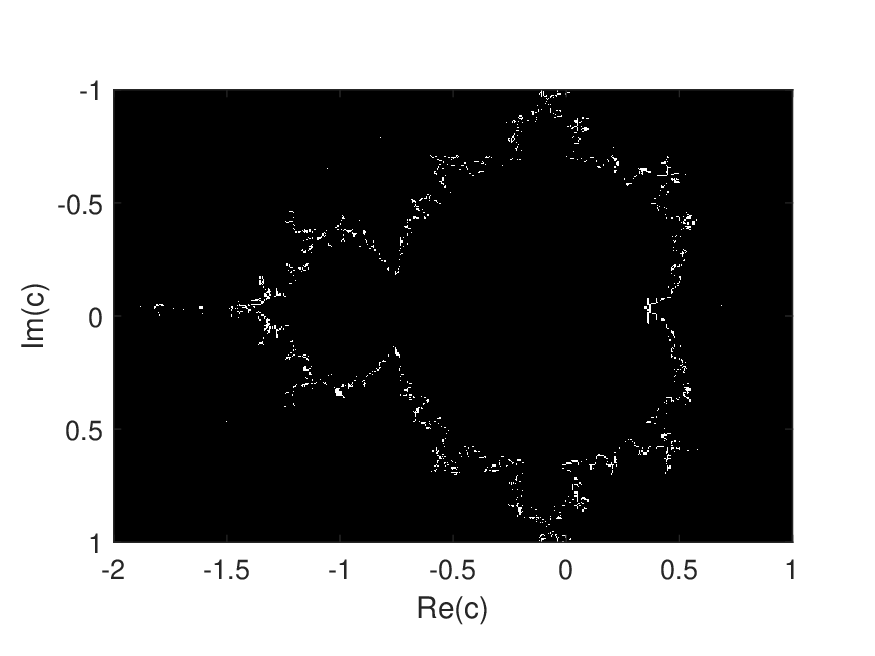}}
\caption{
Classification results on the Mandelbrot set. Top: Performance across different models. Bottom: KNN performance visualization, where black indicates correct classifications and white indicates incorrect classifications.}
\label{fig:violin-mandelbrot}
\end{center}
\vskip -0.2in
\end{figure}

We begin by considering the Mandelbrot set over the described domain above. For the input features, we used the first $I = 4$ nonzero elements of the orbit, resulting in a feature space of dimension $8$. The classification performance on the test set is visualized in Fig.~\ref{fig:violin-mandelbrot} and summarized in the rightmost column of Table~\ref{tab:mandelbrot}. 

The best classification performance was achieved by KNN, followed closely by RF, CART, and MLP, all of which demonstrated strong performance with approximately 95\% test accuracy. The CNN achieved around 92\%, while LSTM and BiLSTM models performed slightly below 90\%. All seven ML methods significantly outperformed the THRESH method.

Additionally, the classification performance of KNN is further illustrated in the bottom panel of Fig.~\ref{fig:violin-mandelbrot}, where black represents correct classifications and white represents errors. We observe that the majority of incorrect classifications occur near the boundary of the Mandelbrot set, a trend that was also evident with the other methods.

\subsection{Julia Set}
\label{sec:results-julia}
\begin{figure}[ht]
\vskip 0.2in
\begin{center}
\centerline{\includegraphics[width=\columnwidth]{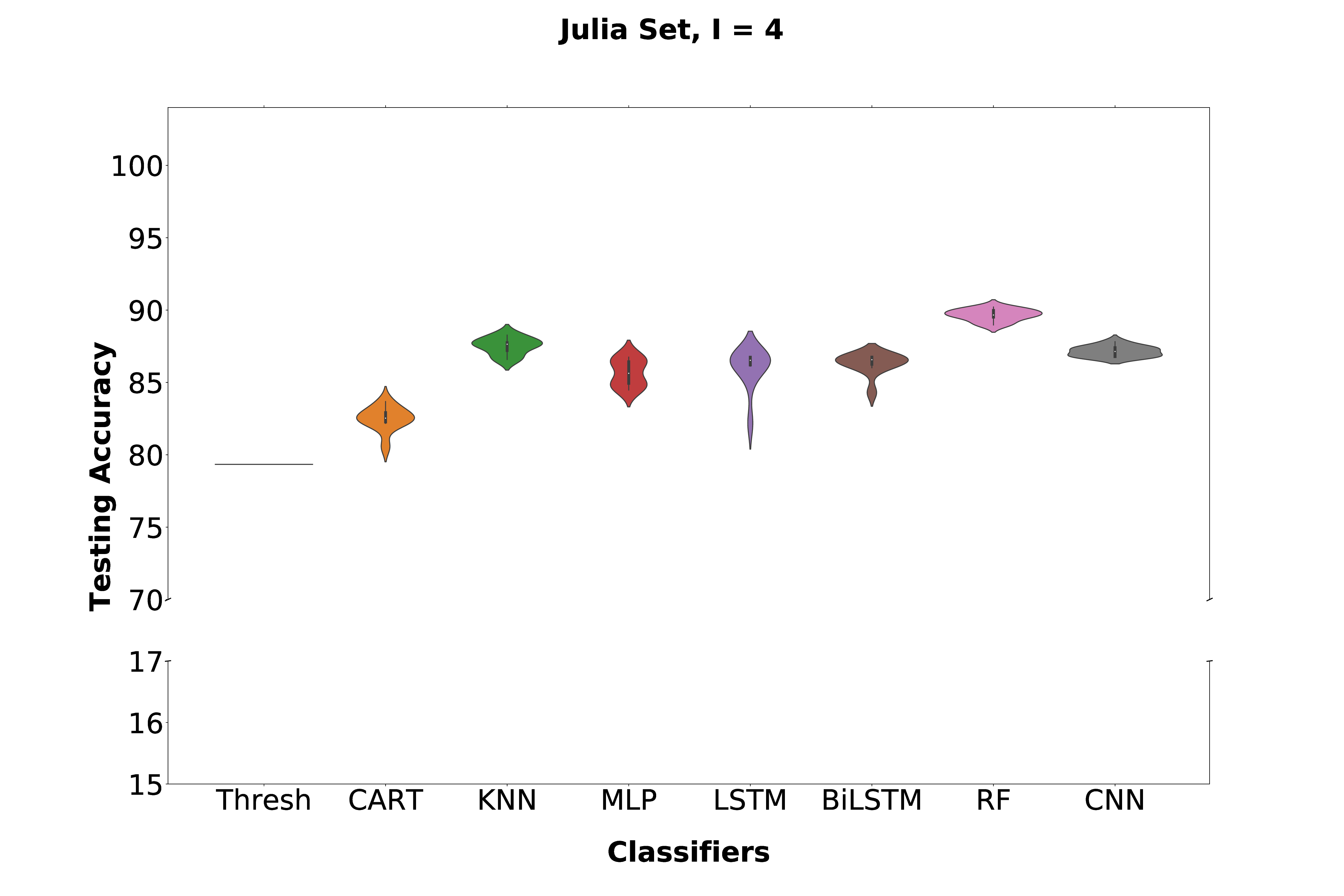}}
\centerline{\includegraphics[width=\columnwidth]{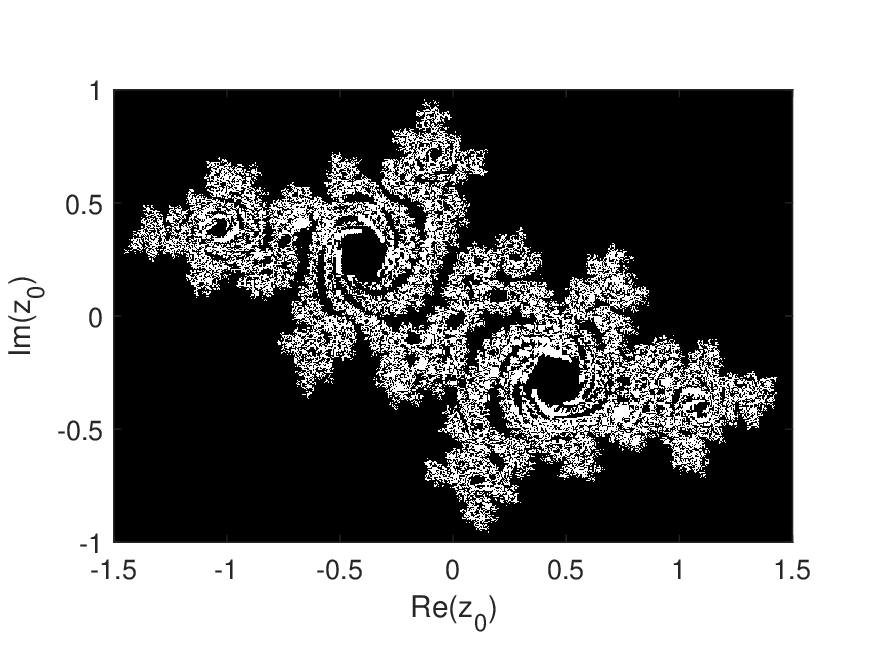}}
\caption{
Classification results on the Julia set. Top: Performance across different models. Bottom: KNN performance visualization, where black indicates correct classifications and white indicates incorrect classifications.}
\label{fig:Violin-Julia}
\end{center}
\vskip -0.2in
\end{figure}

We now study the Julia set with $c = -0.5125 + 0.52129i$. For input features, we used the first $I = 4$ nonzero elements of the orbit, resulting in a feature space of dimension $8$. The test classification performance is visualized in Fig.~\ref{fig:Violin-Julia} and summarized in the rightmost column of Table~\ref{tab:julia}.

The best performance was achieved by RF, with approximately 89.7\%, followed closely by the RNN and CNN, which performed slightly better than LSTM at 86\%, followed by MLP and BiLSTM. Compared to the Mandelbrot set, the overall accuracies for the ML methods are lower, with a significant drop for CART at 82.5\%. This drop is attributed to its greedy approach, which overemphasizes the first iterate (see Appendix \ref{appB} for Gini and $\chi^2$-based feature importance metrics).

The threshold-based method showed a slight increase in accuracy, rising from 78.9\% to 79.3\%. The performance of KNN is visualized in the bottom panel of Fig.~\ref{fig:Violin-Julia}, which partially explains the classification errors. As with the Mandelbrot set, errors predominantly occur near the boundary, but the errors are more dispersed across the domain of study in the Julia set. Interestingly, this behavior does not fully explain why THRESH exhibited no performance degradation, while the ML methods did.

\subsection{Mandelbrot Set Near Boundary}
\label{sec:results-mandelbrot-boundary}

\begin{figure}[ht]
\vskip 0.2in
\begin{center}
\centerline{\includegraphics[width=\columnwidth]{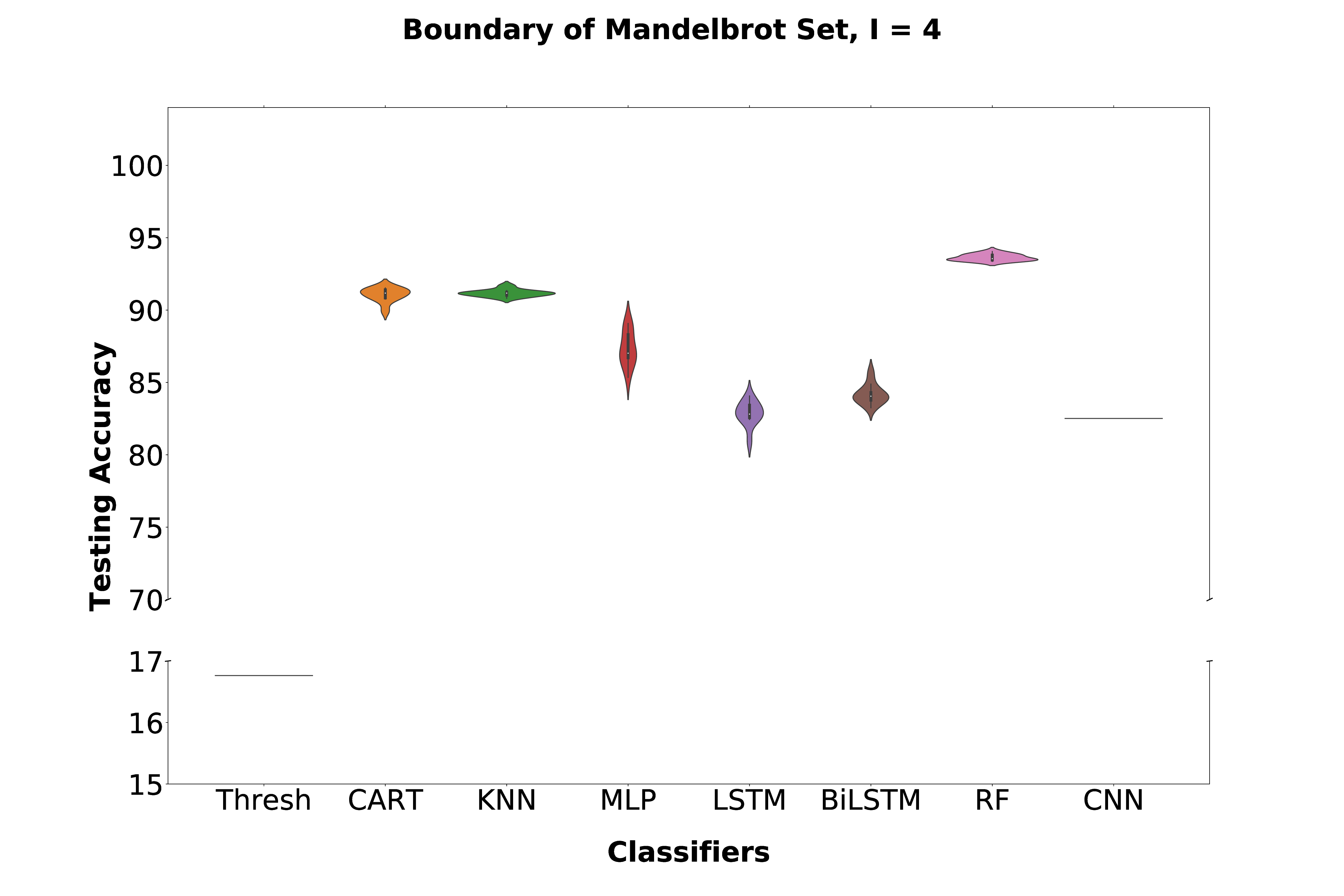}}
\caption{Classification results on the Mandelbrot set close to its boundary.}
\label{fig:violin-boundary-mandelbrot}
\end{center}
\vskip -0.2in
\end{figure}

The results in Sections~\ref{sec:results-mandelbrot} and \ref{sec:results-julia} suggest that classification performance is influenced by the proximity of the initial value to the boundary of the corresponding fractal set. To further investigate this, we employed an approximation method to the boundary of the Mandelbrot set (Section~\ref{sec:boundary}) and trained and tested classifier performance on points near the Mandelbrot boundary. The classification results for this experiment are presented in Fig.~\ref{fig:violin-boundary-mandelbrot} and summarized in the rightmost column of Table~\ref{tab:mandelbrot-boundary}.

An interesting set of results emerges from this experiment. The threshold-based method demonstrated performance worse than random chance, with an accuracy of approximately 17\%. Several models, including MLP, LSTM, BiLSTM, and CNN, collapsed to the trivial classifier, which predicted "unbounded" for all points. This outcome was due to a severe class imbalance in the training data, where 82.5\% of the sampled points were labeled as unbounded. This class imbalance was also reflected in the test set, where the proportion of unbounded orbits was identical, resulting in zero variance across all test runs.

In contrast, CART, KNN, and RF all produced nontrivial classifiers, maintaining strong performance (greater than or equal to 90\%), though slightly lower than the results observed in the original Mandelbrot set experiments (Table~\ref{tab:mandelbrot}). These results suggest interesting avenues for future work, such as modifying the sampling strategy from a uniform distribution to one that is more aware of the boundary, which we discuss in Section~\ref{sec:conclusion}.


Taken together, these results suggest the presence of underlying structures within the orbits of these fractal sets that warrant further exploration and understanding.

\subsection{Learning with Fewer Iterations}
\label{sec:results-learnfewer}

The results presented in Sections~\ref{sec:results-mandelbrot}, \ref{sec:results-julia}, and \ref{sec:results-mandelbrot-boundary} were obtained using the first four iterates of the corresponding fractal orbit ($I = 4$). To study the sensitivity to $I$, Tables~\ref{tab:mandelbrot}, \ref{tab:julia}, and \ref{tab:mandelbrot-boundary} also summarize classification performance for $I = 1, 2, 3$. 

We first observe that all methods generally show an increase in performance as the number of input features increases. However, some methods exhibit more significant improvements, while others maintain relatively consistent performance across values of $I$. For example, Table~\ref{tab:mandelbrot} shows that CART maintains a stable performance across all values of $I$, while BiLSTM increases from approximately 70\% accuracy ($I=1$) to nearly 90\% accuracy when $I=4$.

\textcolor{sam}{RF consistently outperforms CART across most simulation runs, except for the case when $I=1$, suggesting that CART’s relatively weaker performance for $I=2, 3, 4$ is due to model expressiveness rather than limitations inherent to tree-based algorithms. Our 1D CNN results also show consistent performance above chance, competitive with KNN/LSTM on the Julia set, but weaker than RF/KNN on the Mandelbrot set. The CNN model collapses to the trivial classifier, similar to MLP, LSTM, and BiLSTM, when applied to the "fractal zone" sampling described in Section~\ref{sec:results-mandelbrot-boundary}. This suggests that convolution filters capture robust signals when the class boundary is more distinct (e.g., in the Julia set), but fail to identify subtle correlations, as observed in the boundary task. We hypothesize that deeper CNNs or 2-D kernels applied to the complex plane might address this gap.}

\textcolor{sam}{Additionally, we observe that in all cases, the classical THRESH algorithm performs the worst, often by a substantial margin. These calculations highlight that the relative superiority of the ML algorithms is not highly sensitive to the number of iterates used as features. Furthermore, the way some methods' performance changes with different values of $I$ suggests interesting avenues for further exploration, as discussed in Section~\ref{sec:conclusion}.}

\begin{table}[ht!]
\caption{Mandelbrot set classification performance using different numbers of iterates as input variables}
\label{table1}
\label{tab:mandelbrot}
\vskip 0.15in
\begin{center}
\begin{small}
\begin{sc}
\begin{tabular}{lllllr}
\toprule
Method & $I=1$ &  $I=2$ &  $I=3$ &  $I=4$ \\
\midrule
THRESH    
& 32.9 
& 45.4
& 68.5 & 78.9 \\
CART 
& 94.1$\pm$0.4
& 94.3$\pm$0.3
& 94.4$\pm$0.4
& 94.4$\pm$0.4
\\
KNN   
& 95.6$\pm$0.3
& 95.7$\pm$0.3
& 95.7$\pm$0.3
&95.8$\pm$0.3 \\
MLP    
& 91.1$\pm$7.7
& 94.3$\pm$0.4
&  94.5$\pm$0.3
&94.9$\pm$0.3
       \\
LSTM     
& 74.7$\pm$5.4
& 80.1$\pm$2.7
& 81.7$\pm$3.6
& 87.1$\pm$1.5
\\
BiLSTM      
&70.6$\pm$5.1
& 78.1$\pm$4.6
& 82.4$\pm$2.5
& 89.6$\pm$1.8 \\ 
\textcolor{sam}{
RF}
&95.7$\pm$0.3
& 95.6$\pm$0.3
& 95.5$\pm$0.3
& 95.6$\pm$0.2 \\
\textcolor{sam}{CNN}
&87.0$\pm$0.3
&89.0$\pm$0.3
&90.5$\pm$0.3
&92.4$\pm$0.4 \\
\bottomrule
\end{tabular}
\end{sc}
\end{small}
\end{center}
\vskip -0.1in
\end{table}

\begin{table}[ht!]
\caption{Julia set classification performance using different numbers of iterates as input variables}
\label{table2}
\label{tab:julia}
\vskip 0.15in
\begin{center}
\begin{small}
\begin{sc}
\begin{tabular}{lllllr}
\toprule
Method & $I=1$ &  $I=2$ &  $I=3$ &  $I=4$ \\
\midrule
THRESH    
& 20.4
& 39.3
& 55.7
& 79.3 \\
CART 
& 76.1$\pm$0.6
& 80.7$\pm$0.8
& 81.6$\pm$0.7
& 82.5$\pm$0.8
\\
KNN 
& 87.7$\pm$0.5
& 87.6$\pm$0.5
& 87.6$\pm$0.6
& 87.5$\pm$0.6 \\
MLP    
& 85.8$\pm$0.7
& 85.9$\pm$0.7
& 85.7$\pm$0.9
& 85.7$\pm$0.9

       \\
LSTM     
& 80.6$\pm$1.9
& 82.1$\pm$2.8
& 83.6$\pm$1.9
& 86.0$\pm$1.4
\\
BiLSTM      
& 80.8$\pm$2.9
& 83.8$\pm$2.3
& 85.5$\pm$1.3
& 86.3$\pm$0.8\\

\textcolor{sam}{
RF}
&87.6$\pm$0.4
& 88.6$\pm$0.4
& 89.3$\pm$0.4
& 89.7$\pm$0.4 \\
\textcolor{sam}{CNN}
&86.7$\pm$0.1
&87.5$\pm$0.5
&87.4$\pm$0.3
&87.1$\pm$0.3 \\
\bottomrule
\end{tabular}
\end{sc}
\end{small}
\end{center}
\vskip -0.1in
\end{table}

\begin{table}[ht!]
\caption{Mandelbrot boundary classification performance using different numbers of iterates as input variables}
\label{table3}
\label{tab:mandelbrot-boundary}
\vskip 0.15in
\begin{center}
\begin{small}
\begin{sc}
\begin{tabular}{lllllr}
\toprule
Method & $I=1$ &  $I=2$ &  $I=3$ &  $I=4$ \\
\midrule
THRESH    
& 16.2$\pm$0
& 16.4$\pm$0
& 16.4$\pm$0
& 16.8$\pm$0 \\
CART 
& 91.0$\pm$0.2
& 91.2$\pm$0.3
& 91.2$\pm$0.4
& 91.1$\pm$0.5
\\
KNN  
& 91.1$\pm$0.2
& 91.1$\pm$0.2
& 91.2$\pm$0.2
& 91.2$\pm$0.2
 \\
MLP    
& 82.5$\pm$0
& 82.5$\pm$0
& 82.5$\pm$0
& 82.5$\pm$0
       \\
LSTM     
& 82.5$\pm$0
& 82.5$\pm$0
& 82.5$\pm$0
& 82.5$\pm$0
\\
BiLSTM      
& 82.5$\pm$0
& 82.5$\pm$0
& 82.5$\pm$0
& 82.5$\pm$0\\
\textcolor{sam}{
RF}
&92.7$\pm$0.2
& 93.3$\pm$0.2
& 93.6$\pm$0.2
& 93.6$\pm$0.2 \\
\textcolor{sam}{CNN}
&82.5$\pm$0.0
&82.5$\pm$0.0
&82.5$\pm$0.0
&82.5$\pm$0.0 \\
\bottomrule
\end{tabular}
\end{sc}
\end{small}
\end{center}
\vskip -0.1in
\end{table}


\section{\textcolor{sam}{Discussion and Conclusions}}
\label{sec:conclusion}

In this study, we conducted a series of computational experiments to evaluate the effectiveness of various ML models in classifying fractal points based on the first few ($\leq 4$) iterates of their orbits. These experiments were performed on both the Mandelbrot and Julia sets, with particular attention to regions near the boundary of the Mandelbrot set. All evaluated ML methods \textcolor{sam}{(see Table~\ref{tab:models})} outperformed the classical thresholding method traditionally used in fractal analysis.

Remarkably, high classification accuracy was achieved using only four orbit iterates, despite the target variable representing an asymptotic property of the dynamical system. Specifically, the \textcolor{sam}{best performing} models achieved approximately 95\% accuracy on the Mandelbrot set and around 88\% on the Julia set. Among the models tested, KNN \textcolor{sam}{and RF} consistently delivered the highest performance.

\textcolor{sam}{Interestingly, the recurrent architectures—namely LSTM and BiLSTM—did not outperform simpler models such as RF or CNN. We hypothesize that this performance gap is due to the short sequence lengths used in our experiments ($I \leq 4$), which limit the advantages of temporal dependency modeling, a core strength of recurrent neural networks. With such short input sequences, the ability of RNNs to model long-range dependencies is underutilized. Future investigations employing longer orbit sequences or more complex dynamical systems may provide further insight into the conditions under which recurrent architectures can meaningfully surpass simpler alternatives in fractal classification tasks.}

\textcolor{sam}{These observations}, combined with the fact that the classical threshold method performs poorly, suggest that something more interesting may be happening in the first few iterates of the fractal orbits than previously thought. \textcolor{sam}{Indeed, the strong performance of KNN and RF} suggests that, despite the infinitely complex patterns of fractal geometry, there exists some regularity in the iterates of these orbits. We posit the following hypothesis:

\begin{hypothesis}
    \textcolor{sam}{Let $\mathcal{M} \subset \mathbb{C}$ denote the Mandelbrot (or Julia, or other similar fractal) set, defined as the set of parameters $c \in \mathbb{C}$ for which the orbit $(z_n)_{n \geq 0}$, under the iteration
\[
z_{n+1} = z_n^2 + c, \quad z_0 = 0,
\]
remains bounded. Then, for any $\varepsilon > 0$, there exists a compact neighborhood $U \subset \mathbb{C}$ of the boundary $\partial \mathcal{M}$, a positive integer $I \in \mathbb{N}$, and a measurable function $f : \mathbb{R}^I \to \{0,1\}$ such that
\[
\mathbb{P}_{c \sim \mu_U} \left[ f\left( |z_1|, |z_2|, \ldots, |z_I| \right) \neq \mathbf{1}_{\{c \in \mathcal{M}\}} \right] < \varepsilon,
\]
where $z_n = z_n(c)$ is the $n$-th iterate of the orbit, $\mu_U$ denotes the uniform probability measure on $U$, and $\mathbf{1}_{\{c \in \mathcal{M}\}}$ is the indicator function for membership in $\mathcal{M}$. Moreover, there exists a sequence of functions $\{f_I: \mathbb{R}^I \to \{0,1\}\}_{I \in \mathbb{N}}$ such that the corresponding classification error converges rapidly to zero as $I \to \infty$.}
\end{hypothesis}

\textcolor{sam}{
These findings suggest that early orbit iterates encode significant geometric information about the asymptotic behavior of the system. They also point to a promising direction for future research: quantitatively estimating the rate of convergence—presumed to exist—of the classification error as a function of the number of iterates.}

To our knowledge, such properties have not been explored. Furthermore, there are other puzzles arising from the above results: Why do KNN and CART maintain relatively stable performance when using different numbers of iterates, while MLP, LSTM, and BiLSTM show significant improvements (see Section~\ref{sec:results-learnfewer})? \textcolor{sam}{Our results demonstrate that LSTM and BiLSTM were consistently outperformed by RF, CNN, and most of the other methods (except THRESH) on our suite of experiments, with their performance never exceeding 90\% on any task. We attribute this to the short sequence length, where $I \leq 4$. In such cases, temporal dependency modeling offers little benefit, and the extra parameters of a (Bi)LSTM add model variance, thereby penalizing performance.}

\textcolor{sam}{Previous work \cite{boulle2020classification} has demonstrated the effectiveness of deep learning algorithms in classifying chaotic time series. However, these studies typically rely on empirical signals from physical dynamical systems or synthetically generated data based on known attractors. In contrast, our results focus on the classification of rigorously defined mathematical objects, specifically the asymptotic orbit boundedness associated with the Mandelbrot and Julia sets. A promising direction for future research would be to incorporate boundary-aware oversampling techniques near the fractal boundary, which could enhance classifier robustness and potentially uncover additional structural features in these highly complex regions. Such efforts would further contribute to the broader program of applying ML to uncover latent structures in deterministic mathematical systems.}


\textcolor{sam}{While the prior work discussed in the Introduction and Section~\ref{sec:relatedwork} represents significant progress at the intersection of fractal geometry and ML, we wish the reader to observe 3 distinguishing features of our investigation.  First, whereas previous studies primarily focus on symbolic sequences, time series, or empirical dynamical systems, our approach applies ML methods to rigorously defined mathematical fractals—specifically the Mandelbrot and Julia sets—where the ground truth is analytically determined. This perspective enables an investigation of learning behavior in contexts where correctness is defined through precise mathematical properties. Second, many of the referenced works emphasize predictive accuracy and rely on long simulations or extended orbit sequences. By contrast, our focus lies in evaluating whether classifiers can infer \emph{asymptotic boundedness} using only a few initial iterates. This reframes the problem as one of short-time predictive generalization, raising fundamental questions about the information content of early dynamical behavior. Third, we prioritize interpretability over predictive optimization. Rather than tuning models solely for performance, we undertake a comparative analysis across different model families to explore the types of geometric and dynamical features each model captures. In this way, the ML models serve not only as predictors but also as analytical tools for isolating and identifying structural patterns in complex fractal sets.
}

More generally, \textcolor{sam}{our} work contributes to a growing body of \textcolor{sam}{research demonstrating} that ML methods can be applied, somewhat ironically, to pure mathematical objects. Indeed, other \textcolor{sam}{applications} of ML to dynamical systems and partial differential equations have already begun to establish this trend \cite{champion2019data,iten2020discovering,raissi2018deep,raissi2018hidden,raissi2019physics}. Our hope is that ML, as an applied mathematics tool, will increasingly be used to address problems in pure mathematics.



\section*{Data availability}
No data was used for the research described in the article.

\section*{Declaration of competing interest} 
The authors declare that they have no known competing financial interests or personal relationships that could have appeared to influence the work reported in this paper.

\section*{Credit authorship contribution statement}
The manuscript was written with contributions from all authors. All authors have given their approval to the final version of the manuscript.

\textbf{VRT}: Software, Validation, Formal Analysis, Investigation, Writing - Original Draft; \textbf{SFF}: Methodology, Formal Analysis, Writing - Review \& Editing; \textbf{ERMP}: Writing - Review \& Editing, Project Administration, Funding Acquisition; \textbf{HS}: Conceptualization, Supervision, Writing - Review \& Editing.

\begin{acknowledgements}
HS acknowledged support by Khalifa University through a Competitive Internal Research Awards Grant (No.\ 8474000413/CIRA-2021-065) and Research \& Innovation Grants (No.\ 8474000617/RIG-S-2023-031 and No.\ 8474000789/RIG-S-2024-070). \textcolor{sam}{SF acknowledged support from SAFIR Research Institute and Sorbonne Center for Artificial Intelligence, Abu Dhabi. The authors also acknowledge the three independent reviewers for their helpful comments which resulted in improving the quality of the paper.}
\end{acknowledgements}

\appendix

\section{Training Details}
\label{appA}
\textcolor{sam}{
A GitHub repository containing all implementation details necessary to reproduce our results is available at \cite{fractalpaper}. 
Table~\ref{tab:parameters} provides the architecture and key (hyper)parameter values used in the numerical experiments.}

    \begin{table}[htbp!]
        \centering
        \caption{Model Parameters}
        \vskip 0.15in
        \label{tab:parameters}
        \begin{tabular}{cc}
            \hline
            \textbf{\underline{CART Parameter}} & \textbf{\underline{Value}} \\
            {\small OptimizeHyperparameters} & auto\\
            \multirow{2}[0]{*}{\scriptsize HyperparameterOptimizationOptions} & {\scriptsize struct('AcquisitionFunctionName'}\\
            & {\scriptsize ,'expected-improvement-plus')}\\
            \hline
            \textbf{\underline{KNN Parameter}} & \textbf{\underline{Value}} \\
            NumNeighbors & 3\\
            Distance & Euclidean \\
            \hline
            \textbf{\underline{MLP Parameter}} & \textbf{\underline{Value}} \\
            hidden\_layer\_sizes & Fully Connected Layer (7,7,7)\\
            activation           & Logistic Function\\
            solver               & Stochastic Gradient Descent \\
            learning\_rate       & Adaptive Type\\
            max\_iter            & 150\\
            \hline
            \textbf{\underline{LSTM/BiLSTM Parameter}} & \textbf{\underline{Value}} \\
            sequenceInputLayer   & 8\\
            fullyConnectedLayer  & 7\\
            LSTMunits/biLSTMUnits            & 7\\
            fullyConnectedLayer  & 2\\
            Activation Layer     & softmaxLayer\\
            Additional Layer     & classificationLayer\\
            Optimizer            & Stochastic Gradient Descent\\
            MaxEpochs		     & 200\\
            GradientThreshold    & 1\\
            InitialLearnRate     & 0.005\\
            LearnRateSchedule    & piecewise\\
            LearnRateDropPeriod  & 25\\
            LearnRateDropFactor  & 0.2\\
            Epoch                & 200\\
            \hline
            \textbf{\underline{RF Parameter}} & \textbf{\underline{Value}} \\
            numTrees      & 100\\
            Method        & Classification\\
            OOBPrediction & on\\
            OOBVarImp     & on\\
            \hline
            \textbf{\underline{CNN Parameter}} & \textbf{\underline{Value}} \\
            Conv1D                 & Filter = 32\\
            Activation for Conv1D  & ReLU\\
            Dropout on Conv1D      & 0.2\\
            GlobalMaxPooling1D     & on\\
            Dense Layer            & 16\\
            Activation for Conv1D  & ReLU\\
            Dropout on Dense Layer & 0.3\\
            Final Activation Layer & Sigmoid\\
            \hline
        \end{tabular}
    \end{table}

\textcolor{sam}{
Our analysis pipeline began by generating a full dataset of $1,000,000$ observations for each experiment. From this base dataset, disjoint subsets of 10,000 rows were extracted to create ten independent training datasets for each experiment-model combination. Each training dataset was used to train a model, which was then evaluated on a separate test set, also consisting of $1,000,000$ observations, constructed independently.
}


\section{Gini- and $\chi^2$-Based Feature Scores}
\label{appB}

\begin{figure}[ht!]
\vskip 0.2in
\begin{center}
\centerline{\includegraphics[width=\columnwidth]{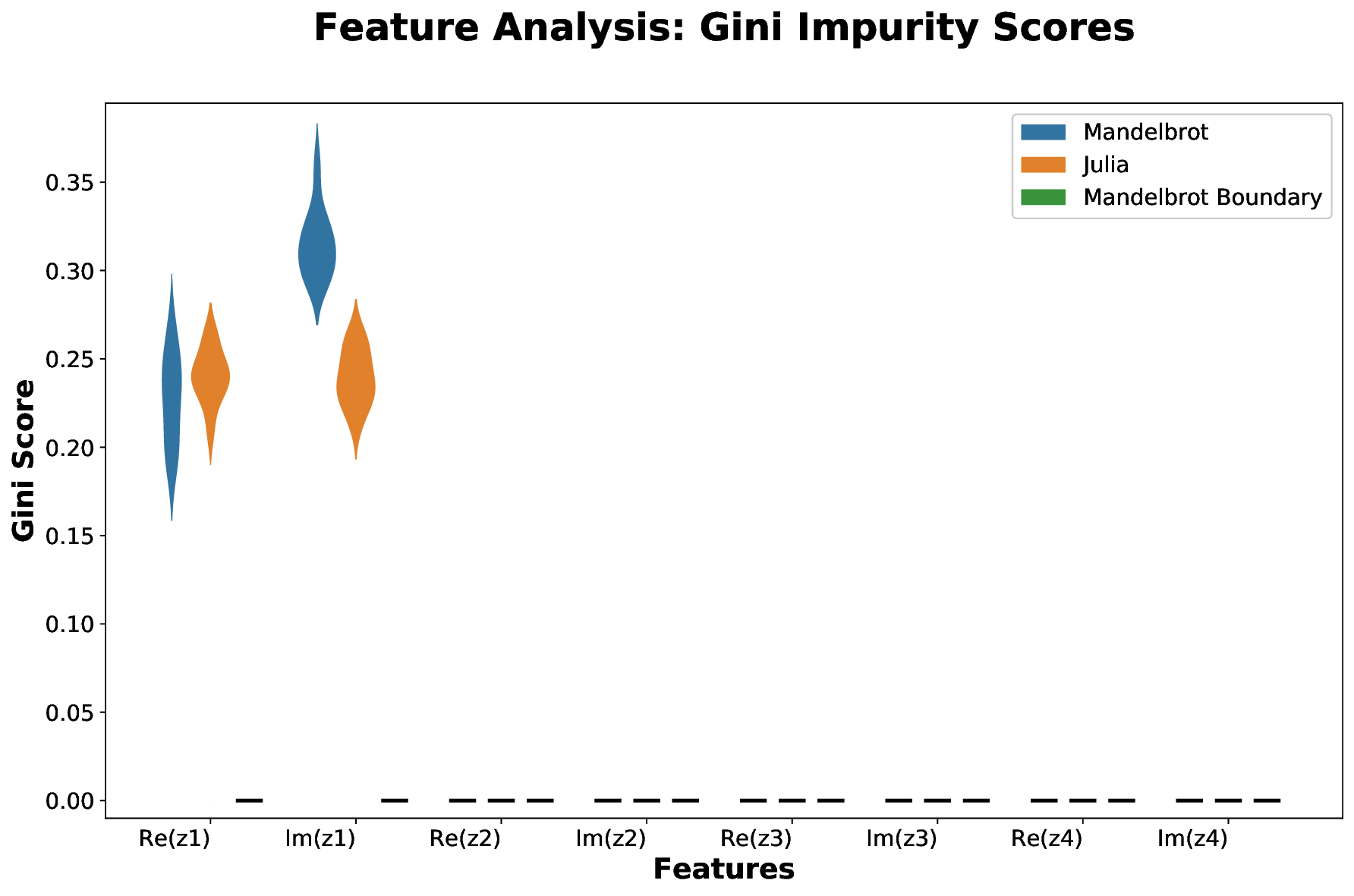}}
\caption{Feature analysis using the Gini index for each feature, showing their ranked importance. The dashed black lines indicate zero values.}
\label{fig:violin-gini}
\end{center}
\vskip -0.2in
\end{figure}

\textcolor{sam}{We further analyzed the features used by CART to diagnose its relatively poor performance, particularly on the Julia set. Over 10 independent simulation runs, we computed the univariate Gini filter scores for the real and imaginary parts of each of the four iterates. The results are presented in Fig.~\ref{fig:violin-gini}, which shows that almost all of CART's splits occur on the first iterate ($\operatorname{Re}(z_1)$ and $\operatorname{Im}(z_1)$), while the Gini scores for $\{\operatorname{Re}(z_2), \dots, \operatorname{Im}(z_4)\}$ are negligible.}

\textcolor{sam}{As discussed in the main text, this indicates that CART exhibits a greedy bias toward early orbit features, which limits its expressiveness by not fully considering the relatively weaker predictors. The fact that all eight of our features are deterministically linked highlights that CART fails to leverage the richer dynamics found in the later iterates.}

\begin{figure}[ht!]
\vskip 0.2in
\begin{center}
\centerline{\includegraphics[width=\columnwidth]{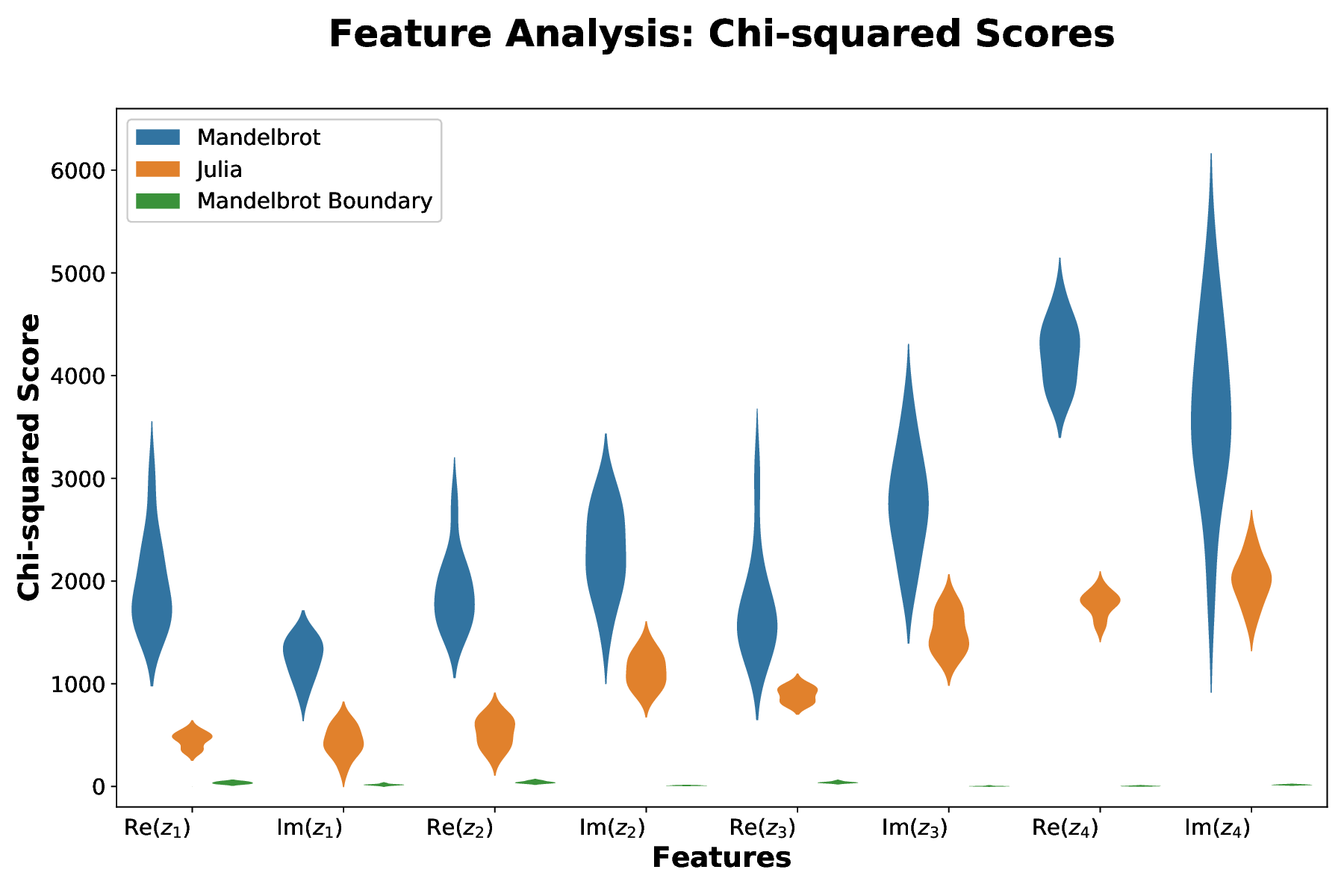}}
\caption{The $\chi^2$-test evaluates the significance of each categorical feature in the model.}
\label{fig:violin-chisq}
\end{center}
\vskip -0.2in
\end{figure}

\textcolor{sam}{In comparison, to compute a purely statistical (model-free) measure of the relevant signal in each feature, we discretized all eight features into three bins and computed the $\chi^2$ statistic for each feature in relation to the output class (bounded/unbounded). The results are shown in Fig.~\ref{fig:violin-chisq}. As expected, there is a general trend for later iterates to exhibit stronger class associations. This observation reinforces our intuition from Figure~\ref{fig:violin-gini} that CART's greedy approach focuses on the first iterate, discarding valuable (albeit more subtle) information from the deeper orbits, which results in relatively weaker performance. More flexible models, such as RF, which were discussed in the main text, can effectively utilize these features and thereby improve classification performance.}

\begin{thebibliography}{10}
	
	\bibitem{gouyet1996physics}
	Jean-Fran{\c{c}}ois Gouyet.
	\newblock {\em Physics and Fractal Structures}.
	\newblock Masson Springer, 1996.
	
	\bibitem{mandelbrot1982fractal}
	Benoit~B Mandelbrot.
	\newblock {\em The Fractal Geometry of Nature}, volume~1.
	\newblock WH Freeman New York, 1982.
	
	\bibitem{kunze2011fractal}
	Herb Kunze, Davide La~Torre, Franklin Mendivil, and Edward~R Vrscay.
	\newblock {\em Fractal-based methods in analysis}.
	\newblock Springer Science \& Business Media, 2011.
	
	\bibitem{katok1995introduction}
	Anatole Katok and Boris Hasselblatt.
	\newblock {\em Introduction to the Modern Theory of Dynamical Systems},
	volume~54 of {\em Encyclopedia of Mathematics and its Applications}.
	\newblock Cambridge University Press, Cambridge, 1995.
	
	\bibitem{devaney1993first}
	Robert~L Devaney, Peter~B Siegel, A~John Mallinckrodt, and Susan McKay.
	\newblock A first course in chaotic dynamical systems: Theory and experiment.
	\newblock {\em Computers in Physics}, 7(4):416--417, 1993.
	
	\bibitem{danca2023mandelbrot}
	Marius-F Danca and Michal Fe{\v{c}}kan.
	\newblock {M}andelbrot set and {J}ulia sets of fractional order.
	\newblock {\em Nonlinear Dynamics}, 111(10):9555--9570, 2023.
	
	\bibitem{kawahira2020zalcman}
	Tomoki Kawahira.
	\newblock Zalcman functions and similarity between the {M}andelbrot set,
	{J}ulia sets, and the tricorn.
	\newblock {\em Analysis and Mathematical Physics}, 10(2):16, 2020.
	
	\bibitem{lei1990similarity}
	Tan Lei.
	\newblock Similarity between the {M}andelbrot set and {J}ulia sets.
	\newblock {\em Communications in mathematical physics}, 134:587--617, 1990.
	
	\bibitem{dobbs2023hausdorff}
	Neil Dobbs, Jacek Graczyk, and Nicolae Mihalache.
	\newblock Hausdorff dimension of {J}ulia sets in the logistic family.
	\newblock {\em Communications in Mathematical Physics}, 399(2):673--716, 2023.
	
	\bibitem{jaksztas2023directional}
	Ludwik Jaksztas.
	\newblock On the directional derivative of the {H}ausdorff dimension of
	quadratic polynomial {J}ulia sets at-2.
	\newblock {\em Advances in Mathematics}, 433:109297, 2023.
	
	\bibitem{chan2016automatic}
	Alan Chan and Jack~A Tuszynski.
	\newblock Automatic prediction of tumour malignancy in breast cancer with
	fractal dimension.
	\newblock {\em Royal Society open science}, 3(12):160558, 2016.
	
	\bibitem{ali2020machine}
	Abder-Rahman Ali, Jingpeng Li, Guang Yang, and Sally~Jane O’Shea.
	\newblock A machine learning approach to automatic detection of irregularity in
	skin lesion border using dermoscopic images.
	\newblock {\em PeerJ Computer Science}, 6:e268, 2020.
	
	\bibitem{hu2020machine}
	Qiao Hu, Yi~Zhou, Shixing Wang, and Futao Wang.
	\newblock Machine learning and fractal theory models for landslide
	susceptibility mapping: Case study from the {J}insha {R}iver {B}asin.
	\newblock {\em Geomorphology}, 351:106975, 2020.
	
	\bibitem{tino2001predicting}
	Peter Tino and Georg Dorffner.
	\newblock Predicting the future of discrete sequences from fractal
	representations of the past.
	\newblock {\em Machine Learning}, 45(2):187--217, 2001.
	
	\bibitem{wu2007schema}
	Ming-Sheng Wu, Jyh-Horng Jeng, and Jer-Guang Hsieh.
	\newblock Schema genetic algorithm for fractal image compression.
	\newblock {\em Engineering Applications of Artificial Intelligence},
	20(4):531--538, 2007.
	
	\bibitem{kataoka2020pre}
	Hirokatsu Kataoka, Kazushige Okayasu, Asato Matsumoto, Eisuke Yamagata, Ryosuke
	Yamada, Nakamasa Inoue, Akio Nakamura, and Yutaka Satoh.
	\newblock Pre-training without natural images.
	\newblock In {\em Proceedings of the Asian Conference on Computer Vision},
	2020.
	
	\bibitem{anderson2022improving}
	Connor Anderson and Ryan Farrell.
	\newblock Improving fractal pre-training.
	\newblock In {\em Proceedings of the IEEE/CVF Winter Conference on Applications
		of Computer Vision}, pages 1300--1309, 2022.
	
	\bibitem{tu2023learning}
	Cheng-Hao Tu, Hong-You Chen, David Carlyn, and Wei-Lun Chao.
	\newblock Learning fractals by gradient descent.
	\newblock In {\em Proceedings of the AAAI Conference on Artificial
		Intelligence}, volume~37, pages 2456--2464, 2023.
	
	\bibitem{mastersthesis}
	Peter Bloem.
	\newblock Machine learning and fractal geometry.
	\newblock Master's thesis, University of Amsterdam, Amsterdam, 2010.
	
	\bibitem{chatel}
	Gr\'egory Ch\^{a}tel.
	\newblock What do deep neural networks understand of fractals?, 2017.
	
	\bibitem{mandelbrot1983quadratic}
	Benoit~B Mandelbrot.
	\newblock On the quadratic mapping $z\to z^2-\mu$ for complex $\mu$ and $z$:
	the fractal structure of its $\mathcal{M}$ set, and scaling.
	\newblock {\em Physica D: Nonlinear Phenomena}, 7(1-3):224--239, 1983.
	
	\bibitem{blanchard1986disconnected}
	Paul Blanchard.
	\newblock Disconnected {J}ulia sets.
	\newblock In {\em Chaotic Dynamics and Fractals}, pages 181--201. Elsevier,
	1986.
	
	\bibitem{liu2015clinical}
	Zitao Liu and Milos Hauskrecht.
	\newblock Clinical time series prediction: Toward a hierarchical dynamical
	system framework.
	\newblock {\em Artificial intelligence in medicine}, 65(1):5--18, 2015.
	
	\bibitem{castillo1995intelligent}
	Oscar Castillo and Patricia Melin.
	\newblock An intelligent system for financial time series prediction combining
	dynamical systems theory, fractal theory, and statistical methods.
	\newblock In {\em Proceedings of 1995 Conference on Computational Intelligence
		for Financial Engineering (CIFEr)}, pages 151--155. IEEE, 1995.
	
	\bibitem{boulle2020classification}
	Nicolas Boull{\'e}, Vassilios Dallas, Yuji Nakatsukasa, and D~Samaddar.
	\newblock Classification of chaotic time series with deep learning.
	\newblock {\em Physica D: Nonlinear Phenomena}, 403:132261, 2020.
	
	\bibitem{ismail2019deep}
	Hassan Ismail~Fawaz, Germain Forestier, Jonathan Weber, Lhassane Idoumghar, and
	Pierre-Alain Muller.
	\newblock Deep learning for time series classification: a review.
	\newblock {\em Data Mining and Knowledge Discovery}, 33(4):917--963, 2019.
	
	\bibitem{ismail2020inceptiontime}
	Hassan Ismail~Fawaz, Benjamin Lucas, Germain Forestier, Charlotte Pelletier,
	Daniel~F Schmidt, Jonathan Weber, Geoffrey~I Webb, Lhassane Idoumghar,
	Pierre-Alain Muller, and Fran{\c{c}}ois Petitjean.
	\newblock Inceptiontime: Finding alexnet for time series classification.
	\newblock {\em Data Mining and Knowledge Discovery}, 34(6):1936--1962, 2020.
	
	\bibitem{grohs2023proof}
	Philipp Grohs, Fabian Hornung, Arnulf Jentzen, and Philippe von Wurstemberger.
	\newblock {\em A Proof that Artificial Neural Networks Overcome the Curse of
		Dimensionality in the Numerical Approximation of Black--Scholes Partial
		Differential Equations}, volume 284 of {\em Memoirs of the American
		Mathematical Society}.
	\newblock American Mathematical Society, 2023.
	
	\bibitem{han2018solving}
	Jiequn Han, Arnulf Jentzen, and E~Weinan.
	\newblock Solving high-dimensional partial differential equations using deep
	learning.
	\newblock {\em Proceedings of the National Academy of Sciences},
	115(34):8505--8510, 2018.
	
	\bibitem{michoski2020solving}
	Craig Michoski, Milo{\v{s}} Milosavljevi{\'c}, Todd Oliver, and David~R Hatch.
	\newblock Solving differential equations using deep neural networks.
	\newblock {\em Neurocomputing}, 399:193--212, 2020.
	
	\bibitem{putri2024deep}
	Endah~RM Putri, Muhammad~L Shahab, Mohammad Iqbal, Imam Mukhlash, Amirul Hakam,
	Lutfi Mardianto, and Hadi Susanto.
	\newblock {A deep-genetic algorithm (deep-GA) approach for high-dimensional
		nonlinear parabolic partial differential equations}.
	\newblock {\em Computers \& Mathematics with Applications}, 154:120--127, 2024.
	
	\bibitem{raissi2018deep}
	Maziar Raissi.
	\newblock Deep hidden physics models: Deep learning of nonlinear partial
	differential equations.
	\newblock {\em The Journal of Machine Learning Research}, 19(1):932--955, 2018.
	
	\bibitem{raissi2018hidden}
	Maziar Raissi and George~Em Karniadakis.
	\newblock Hidden physics models: Machine learning of nonlinear partial
	differential equations.
	\newblock {\em Journal of Computational Physics}, 357:125--141, 2018.
	
	\bibitem{raissi2019physics}
	Maziar Raissi, Paris Perdikaris, and George~E Karniadakis.
	\newblock Physics-informed neural networks: A deep learning framework for
	solving forward and inverse problems involving nonlinear partial differential
	equations.
	\newblock {\em Journal of Computational Physics}, 378:686--707, 2019.
	
	\bibitem{shahab2024neural}
	Muhammad~Luthfi Shahab and Hadi Susanto.
	\newblock Neural networks for bifurcation and linear stability analysis of
	steady states in partial differential equations.
	\newblock {\em Applied Mathematics and Computation}, 483:128985, 2024.
	
	\bibitem{shahab2025corrigendum}
	Muhammad~Luthfi Shahab and Hadi Susanto.
	\newblock {Corrigendum to “Neural networks for bifurcation and linear
		stability analysis of steady states in partial differential
		equations”[Appl. Math. Comput. 483 (2024) 128985]}.
	\newblock {\em Applied Mathematics and Computation}, 495:129319, 2025.
	
	\bibitem{shahab2025neural}
	Muhammad~Luthfi Shahab, Fidya~Almira Suheri, Rudy Kusdiantara, and Hadi
	Susanto.
	\newblock Neural networks for high-dimensional solutions and snaking
	bifurcations in nonlinear lattices.
	\newblock Manuscript under review, 2025.
	
	\bibitem{wang2021understanding}
	Sifan Wang, Yujun Teng, and Paris Perdikaris.
	\newblock Understanding and mitigating gradient flow pathologies in
	physics-informed neural networks.
	\newblock {\em SIAM Journal on Scientific Computing}, 43(5):A3055--A3081, 2021.
	
	\bibitem{jagtap2020extended}
	Ameya~D Jagtap and George~Em Karniadakis.
	\newblock {Extended physics-informed neural networks (XPINNs): A generalized
		space-time domain decomposition based deep learning framework for nonlinear
		partial differential equations}.
	\newblock {\em Communications in Computational Physics}, 28(5), 2020.
	
	\bibitem{meng2020composite}
	Xuhui Meng and George~Em Karniadakis.
	\newblock {A composite neural network that learns from multi-fidelity data:
		Application to function approximation and inverse PDE problems}.
	\newblock {\em Journal of Computational Physics}, 401:109020, 2020.
	
	\bibitem{daneker2024transfer}
	Mitchell Daneker, Shengze Cai, Ying Qian, Eric Myzelev, Arsh Kumbhat, He~Li,
	and Lu~Lu.
	\newblock Transfer learning on physics-informed neural networks for tracking
	the hemodynamics in the evolving false lumen of dissected aorta.
	\newblock {\em Nexus}, 1(2), 2024.
	
	\bibitem{robinson2022physics}
	Haakon Robinson, Suraj Pawar, Adil Rasheed, and Omer San.
	\newblock Physics guided neural networks for modelling of non-linear dynamics.
	\newblock {\em Neural Networks}, 154:333--345, 2022.
	
	\bibitem{zhang2024zero}
	Yuanzhao Zhang and William Gilpin.
	\newblock Zero-shot forecasting of chaotic systems.
	\newblock {\em arXiv preprint arXiv:2409.15771}, 2024.
	
	\bibitem{brunton2016discovering}
	Steven~L Brunton, Joshua~L Proctor, and J~Nathan Kutz.
	\newblock Discovering governing equations from data by sparse identification of
	nonlinear dynamical systems.
	\newblock {\em Proceedings of the national academy of sciences},
	113(15):3932--3937, 2016.
	
	\bibitem{champion2019data}
	Kathleen Champion, Bethany Lusch, J~Nathan Kutz, and Steven~L Brunton.
	\newblock Data-driven discovery of coordinates and governing equations.
	\newblock {\em Proceedings of the National Academy of Sciences},
	116(45):22445--22451, 2019.
	
	\bibitem{gilpin2021chaos}
	William Gilpin.
	\newblock Chaos as an interpretable benchmark for forecasting and data-driven
	modelling.
	\newblock {\em arXiv preprint arXiv:2110.05266}, 2021.
	
	\bibitem{fractalpaper}
	V.R.\ Tjahjono.
	\newblock Scripts for {M}andelbrot and {J}ulia paper.
	\newblock \url{https://github.com/venansiusrt/fractal-paper}, 2025.
	\newblock Accessed: 2025-06-13.
	
	\bibitem{iten2020discovering}
	Raban Iten, Tony Metger, Henrik Wilming, L{\'\i}dia Del~Rio, and Renato Renner.
	\newblock Discovering physical concepts with neural networks.
	\newblock {\em Physical Review Letters}, 124(1):010508, 2020.
	
\end{thebibliography}
\end{document}